\documentclass[3p,times]{elsarticle}

\usepackage{ecrc}

\volume{00}

\firstpage{1}

\journalname{High Energy Density Physics}

\runauth{V. Dwarkadas et al.}

\jid{hedp}

\jnltitlelogo{High Energy Density Physics}

\CopyrightLine{2011}{Published by Elsevier Ltd.}

\usepackage{amssymb}
\usepackage{amsthm}
\usepackage{amsmath}

\usepackage[figuresright]{rotating}

\def\go{\mathrel{\raise.3ex\hbox{$>$}\mkern-14mu
             \lower0.6ex\hbox{$\sim$}}}
\def\lo{\mathrel{\raise.3ex\hbox{$<$}\mkern-14mu
             \lower0.6ex\hbox{$\sim$}}}
\def\lsim{\raise0.3ex\hbox{$<$}\kern-0.75em{\lower0.65ex\hbox{$\sim$}}}
\def\gsim{\raise0.3ex\hbox{$>$}\kern-0.75em{\lower0.65ex\hbox{$\sim$}}}

\begin{document}

\newcommand{\vper}{\mbox{${v_{\perp}}$}}
\newcommand{\vpar}{\mbox{${v_{\parallel}}$}}
\newcommand{\uper}{\mbox{${u_{\perp}}$}}
\newcommand{\vperout}{\mbox{${{v_{\perp}}_{o}}$}}
\newcommand{\uperout}{\mbox{${{u_{\perp}}_{o}}$}}
\newcommand{\vperin}{\mbox{${{v_{\perp}}_{i}}$}}
\newcommand{\uperin}{\mbox{${{u_{\perp}}_{i}}$}}
\newcommand{\upar}{\mbox{${u_{\parallel}}$}}
\newcommand{\uparout}{\mbox{${{u_{\parallel}}_{o}}$}}
\newcommand{\vparout}{\mbox{${{v_{\parallel}}_{o}}$}}
\newcommand{\uparin}{\mbox{${{u_{\parallel}}_{i}}$}}
\newcommand{\vparin}{\mbox{${{v_{\parallel}}_{i}}$}}
\newcommand{\dout}{\mbox{${\rho}_{o}$}}
\newcommand{\din}{\mbox{${\rho}_{i}$}}
\newcommand{\da}{\mbox{${\rho}_{1}$}}
\newcommand{\mfast}{\mbox{$\dot{M}_{f}$}}
\newcommand{\mslow}{\mbox{$\dot{M}_{a}$}}
\newcommand{\beqn}{\begin{eqnarray}}
\newcommand{\eeqn}{\end{eqnarray}}
\newcommand{\be}{\begin{equation}}
\newcommand{\ee}{\end{equation}}
\newcommand{\noi}{\noindent}
\newcommand{\ftheta}{\mbox{$f(\theta)$}}
\newcommand{\gtheta}{\mbox{$g(\theta)$}}
\newcommand{\ltheta}{\mbox{$L(\theta)$}}
\newcommand{\stheta}{\mbox{$S(\theta)$}}
\newcommand{\utheta}{\mbox{$U(\theta)$}}
\newcommand{\xitheta}{\mbox{$\xi(\theta)$}}
\newcommand{\vs}{\mbox{${v_{s}}$}}
\newcommand{\ro}{\mbox{${R_{0}}$}}
\newcommand{\pa}{\mbox{${P_{1}}$}}
\newcommand{\va}{\mbox{${v_{a}}$}}
\newcommand{\vo}{\mbox{${v_{o}}$}}
\newcommand{\vp}{\mbox{${v_{p}}$}}
\newcommand{\vw}{\mbox{${v_{w}}$}}
\newcommand{\vf}{\mbox{${v_{f}}$}}
\newcommand{\lprime}{\mbox{${L^{\prime}}$}}
\newcommand{\uprime}{\mbox{${U^{\prime}}$}}
\newcommand{\sprime}{\mbox{${S^{\prime}}$}}
\newcommand{\xiprime}{\mbox{${{\xi}^{\prime}}$}}
\newcommand{\mdot}{\mbox{$\dot{M}$}}
\newcommand{\msun}{\mbox{$M_{\odot}$}}
\newcommand{\yr}{\mbox{${\rm yr}^{-1}$}}
\newcommand{\kms}{\mbox{${\rm km} \;{\rm s}^{-1}$}}
\newcommand{\lambdav}{\mbox{${\lambda}_{v}$}}
\newcommand{\lequ}{\mbox{${L_{eq}}$}}
\newcommand{\eqpratio}{\mbox{${R_{eq}/R_{p}}$}}
\newcommand{\ra}{\mbox{${r_{o}}$}}
\newcommand{\bfig}{\begin{figure}[h]}
\newcommand{\efig}{\end{figure}}
\newcommand{\tone}{\mbox{${t_{1}}$}}
\newcommand{\done}{\mbox{${{\rho}_{1}}$}}
\newcommand{\dsn}{\mbox{${\rho}_{SN}$}}
\newcommand{\dzero}{\mbox{${\rho}_{0}$}}
\newcommand{\ve}{\mbox{${v}_{e}$}}
\newcommand{\vej}{\mbox{${v}_{ej}$}}
\newcommand{\Mch}{\mbox{${M}_{ch}$}}
\newcommand{\mej}{\mbox{${M}_{e}$}}
\newcommand{\Mst}{\mbox{${M}_{ST}$}}
\newcommand{\dam}{\mbox{${\rho}_{am}$}}
\newcommand{\Rst}{\mbox{${R}_{ST}$}}
\newcommand{\Vst}{\mbox{${V}_{ST}$}}
\newcommand{\Tst}{\mbox{${T}_{ST}$}}
\newcommand{\no}{\mbox{${n}_{0}$}}
\newcommand{\Efif}{\mbox{${E}_{51}$}}
\newcommand{\rsh}{\mbox{${R}_{sh}$}}
\newcommand{\msh}{\mbox{${M}_{sh}$}}
\newcommand{\vsh}{\mbox{${V}_{sh}$}}
\newcommand{\vrev}{\mbox{${v}_{rev}$}}
\newcommand{\rpr}{\mbox{${R}^{\prime}$}}
\newcommand{\mpr}{\mbox{${M}^{\prime}$}}
\newcommand{\vpr}{\mbox{${V}^{\prime}$}}
\newcommand{\tpr}{\mbox{${t}^{\prime}$}}
\newcommand{\cone}{\mbox{${c}_{1}$}}
\newcommand{\ctwo}{\mbox{${c}_{2}$}}
\newcommand{\cthree}{\mbox{${c}_{3}$}}
\newcommand{\cfour}{\mbox{${c}_{4}$}}
\newcommand{\Te}{\mbox{${T}_{e}$}}
\newcommand{\Ti}{\mbox{${T}_{i}$}}
\newcommand{\Ha}{\mbox{${H}_{\alpha}$}}
\newcommand{\Rprime}{\mbox{${R}^{\prime}$}}
\newcommand{\Vprime}{\mbox{${V}^{\prime}$}}
\newcommand{\Tprime}{\mbox{${T}^{\prime}$}}
\newcommand{\Mprime}{\mbox{${M}^{\prime}$}}
\newcommand{\rprime}{\mbox{${r}^{\prime}$}}
\newcommand{\rfprime}{\mbox{${r}_f^{\prime}$}}
\newcommand{\vprime}{\mbox{${v}^{\prime}$}}
\newcommand{\tprime}{\mbox{${t}^{\prime}$}}
\newcommand{\mprime}{\mbox{${m}^{\prime}$}}
\newcommand{\Me}{\mbox{${M}_{e}$}}
\newcommand{\nh}{\mbox{${n}_{H}$}}
\newcommand{\rr}{\mbox{${R}_{2}$}}
\newcommand{\rf}{\mbox{${R}_{1}$}}
\newcommand{\vtwo}{\mbox{${V}_{2}$}}
\newcommand{\vout}{\mbox{${V}_{1}$}}
\newcommand{\dshell}{\mbox{${{\rho}_{sh}}$}}
\newcommand{\dwind}{\mbox{${{\rho}_{w}}$}}
\newcommand{\dslow}{\mbox{${{\rho}_{s}}$}}
\newcommand{\dfast}{\mbox{${{\rho}_{f}}$}}
\newcommand{\vfast}{\mbox{${v}_{f}$}}
\newcommand{\vslow}{\mbox{${v}_{s}$}}
\newcommand{\cc}{\mbox{${\rm cm}^{-3}$}}
\newcommand{\apj}{\mbox{ApJ}}
\newcommand{\apjl}{\mbox{ApJL}}
\newcommand{\apjs}{\mbox{ApJS}}
\newcommand{\aj}{\mbox{AJ}}
\newcommand{\araa}{\mbox{ARAA}}
\newcommand{\nat}{\mbox{Nature}}
\newcommand{\aap}{\mbox{AA}}
\newcommand{\gca}{\mbox{GeCoA}}
\newcommand{\pasp}{\mbox{PASP}}
\newcommand{\mnras}{\mbox{MNRAS}}
\newcommand{\apss}{\mbox{ApSS}}

\begin{frontmatter}



\dochead{}

\title{Simulated X-ray Spectra From Ionized Wind-Blown Nebulae around Massive
  Stars}


\author[vvd]{V. V. Dwarkadas\corref{cor1}}
\ead{vikram@oddjob.uchicago.edu}

\author[dr]{D. L. Rosenberg}
\ead{duaner@ucar.edu}

\address[vvd]{Department of Astronomy and astrophysics, Univ of Chicago, TAAC 55, Chicago, IL 60637}

\address[dr]{National Center for Atmospheric Research, Boulder, Colorado}

\begin{abstract}

Using an ionization gasdynamics code, we simulate a model of the
wind-blown bubble around a 40 $\msun$ star. We use this to compute the
X-ray spectra from the bubble, which can be directly compared to
observations. We outline our methods and techniques for these
computations, and contrast them with previous calculations. Our
simulated X-ray spectra compare reasonably well with observed spectra
of Wolf-Rayet bubbles. They suggest that X-ray nebulae around massive
stars may not be easily detectable, consistent with observations.

\end{abstract}


\end{frontmatter}


\section{Introduction}
\label{sec:intro}
Massive stars ($> 8\msun$) lose mass throughout their lifetime, via
winds and eruptions, before ending their lives in a cataclysmic
supernova (SN) explosion. The interaction of this material with the
surrounding medium creates vast wind-blown cavities surrounded by a
dense shell, referred to as wind-blown bubbles. Simultaneously,
radiation from the star can ionize the surrounding medium. As the star
evolves through various stages, the mass-loss parameters, and the
ionizing flux or number of ionizing photons, will change, affecting
the structure of the bubble. When the star finally explodes, the
resulting SN shock wave will expand within the bubble, and the
dynamics and kinematics of the shock wave will depend on the bubble
parameters \cite{dwarkadas2005}. Similarly, the relativistic blast
waves associated with gamma-ray bursts (GRBs) are expected to expand
within wind bubbles surrounding Wolf-Rayet (W-R) stars. Thus, it is
important to understand the structure of the bubble, as it influences
the evolution and emission from SNe and GRBs.

The structure of wind bubbles was identified by Weaver et
al.~\cite{weaver77}. Proceeding radially outward from the star, 4
regions are delineated: (1) a freely expanding wind region, (2) a
shocked wind region, (3) the shocked ambient region, forming a thin
dense shell, and (4) the unshocked ambient medium. An inner, `wind
termination' shock separates region (1) and (2), a contact
discontinuity, regions (2) and (3) and an outer, generally radiative
shock, regions (3) and (4).

In order to understand the structure of wind-bubbles accurately,
models must take into account both the gasdynamics and ionization from
the star. Early models that explored the evolution of massive star
surroundings \cite{glm96, gml96} had some ionization built in, mainly
centered around the Stromgren sphere approximation. They did not model
the main-sequence (MS) stage in multi-dimensions. Similar limitations
were included in the models of \cite{vanmarleetal05,
  vanmarleetal06}. A somewhat better treatment of ionization
properties was included in \cite{fhy03, fhy06}. Dwarkadas
\cite{dwarkadas07a, dwarkadas07c, dwarkadas08} considered the entire
evolution in multi-dimensions, and studied the turbulence in the
interior, but did not include any ionization in his calculations. 3D
simulations carried out by \cite{vanmarleetal11} also do not include
ionization. A more complete treatment of both the ionization plus the
gasdynamics has been included in the work of Arthur and her group
\cite{arthur07, arthur09, ta11}.

In this paper we outline a code, AVATAR, that couples the ionization
to the gasdynamics, and use it to study the evolution of wind bubbles
around a 40 $\msun$ star.  We use a tried and tested method, with the
goal of having a reasonably accurate description of the wind-blown
bubble that is consistent, and at least equivalent to, if not better
than, what other groups have done in the recent past. Our purpose is
to have a numerical description of the bubble using which we can
compute the X-ray emission. Using our code we compute the structure
and evolution of the wind-blown bubble, study the dynamics,
hydrodynamics and kinematics, the formation of instabilities, growth
of small scale structure, and the onset of turbulence.

Much of the volume of the wind bubble is occupied by the
high-pressure, low density shocked wind. Wind velocities of order
1000-2000 km s$^{-1}$ for O, B and Wolf-Rayet (W-R) stars should
conceivably result in post-shock temperatures in the shocked wind of
10$^7$ - 10$^8$ K. It has been assumed in the literature, perhaps
rather naively, that the nebulae should therefore be visible as
regions of diffuse X-ray emission. However, despite extensive searches
around WR and main-sequence stars with Chandra and XMM, diffuse X-ray
emission has been detected in only few cases \cite{chuetal03a,
  chuetal03b, wriggeetal2005}, with observed X-ray luminosities 10-100
times smaller than would be expected given the ``expected''
temperatures and the size of the region. The inferred temperatures for
bubbles around two W-R stars, NGC 6888 and S308, are a few times
10$^6$ K, again lower than expected given the wind
velocities. Recently, Zhekov \& Park \cite{zp11} have found that about
10\% of the flux in 6888 arises from higher temperature plasma ($>$ 2
keV).

It appears that either some of the a-priori assumptions must be
incorrect, or perhaps that the thermal energy is converted to some
other form of energy. Turbulence has been suggested as one answer.
Dwarkadas \cite{dwarkadas08} found, from 2D simulations, that only a
small percentage of the energy went into turbulent motions, except in
the very last stages of W-R evolution, where the energy in non-radial
motions was about 15-20\% of that in the radial flow.  Thermal
conduction is another suggestion, but given the low densities within
the bubble, it seems quite unlikely. This has been quantitatively
shown by \cite{ta11}. In order to address questions regarding the
X-ray emission, we compute X-ray spectra from our simulations, using
the ISIS package \cite{houck2000}.

In this first paper, our intention is to outline our work and
demonstrate the viability of the techniques involved to compute the
X-ray spectra from ionized wind-blown bubbles. We show initial
moderate resolution simulations that demonstrate, given the
limitations of our method, our hydrodynamic results are consistent,
and in some ways better, than others that have been used to compute
the hydrodynamics of wind-bubbles around massive stars. We then
outline our technique for calculating the X-ray emission, and show how
it improves on the work done by other groups. The exciting fact that
our simulated spectra seem to be in reasonable agreement with
observations suggests that we are on the correct track. We will follow
this work with higher resolution simulations and more detailed X-ray
calculations in future papers.

This paper proceeds as follows: In \S 2 we describe the
photoionization code, and in \S 3 the numerical methods. These are
then applied to modelling the medium around a 40 $\msun$ star in \S 4,
and X-ray spectra computed from the models in \S 5. \S 6 summarizes
our work and outlines future development.

\section{AVATAR Code}

We have made significant improvements to an earlier version of this
code \cite{r95}. AVATAR contains a self-consistent method for
computing the effects of photo-ionization on circumstellar gas
dynamics. The effects of geometrical dilution and column absorption of
radiation are considered. The gasdynamic algorithm makes use of a
multidimensional covariant implementation of well established Eulerian
finite difference algorithms. A second-order (van Leer) monotonic
transport algorithm is used for the advection of total mass and the
neutral component, and a third order PPA (piecewise parabolic
algorithm) scheme is available.  Tabulated functions are used to
compute the collisional ionization rate and cooling function. Shocks
are treated using an artificial viscosity. Grid expansion, which may
be required in order to study the flow over distance scales spanning
several orders of magnitude, is incorporated.  The method operator
splits the contribution due to photoionization effects from the usual
gas dynamics, and utilizes a backward-Euler scheme together with a
Newton-Raphson iteration procedure for achieving a solution. The
algorithm incorporates a simplified model of the photo-ionization
source, computes the fractional ionization of hydrogen due to the
photo-ionizing flux and recombination, and determines
self-consistently the energy balance due to ionization, photo-heating
and radiative cooling. In this, our method is superior to that of
other calculations \cite{fhy03, fhy06, glm96, gml96, vanmarleetal05,
  vanmarleetal06} who use the on-the-spot approximation and do not
take recombination into account. It is comparable to the method used
by \cite{ta11}.

\section{Hydrodynamic Models} We  apply the AVATAR code to the computation of 
the wind-blown bubble around a 40 $\msun$ star. The simulations are
carried out on an r-$\theta$ grid in spherical co-ordinates. Runs with
varying resolution were computed, with the largest being 600 radial by
400 angular zones.  The stellar parameters for the 40 $\msun$ star are
adapted from \cite{vanmarleetal05}. The star evolves from the MS stage
into a red-supergiant (RSG), and finally ends its life as a W-R star
before it explodes. Although realistically the stellar parameters are
continuously changing, this would make it extremely time consuming to
calculate the ionization integrals at every timestep. Therefore the
stellar parameters in each phase are assumed constant, as suggested by
\cite{vanmarleetal05}. In this way the integrals need to be computed
only once per stage. In future we will modify this so that we can take
the changing mass-loss rates into account, since these are important
to study the turbulence and instabilities within the bubble.

\begin{figure}
\includegraphics[scale=0.9]{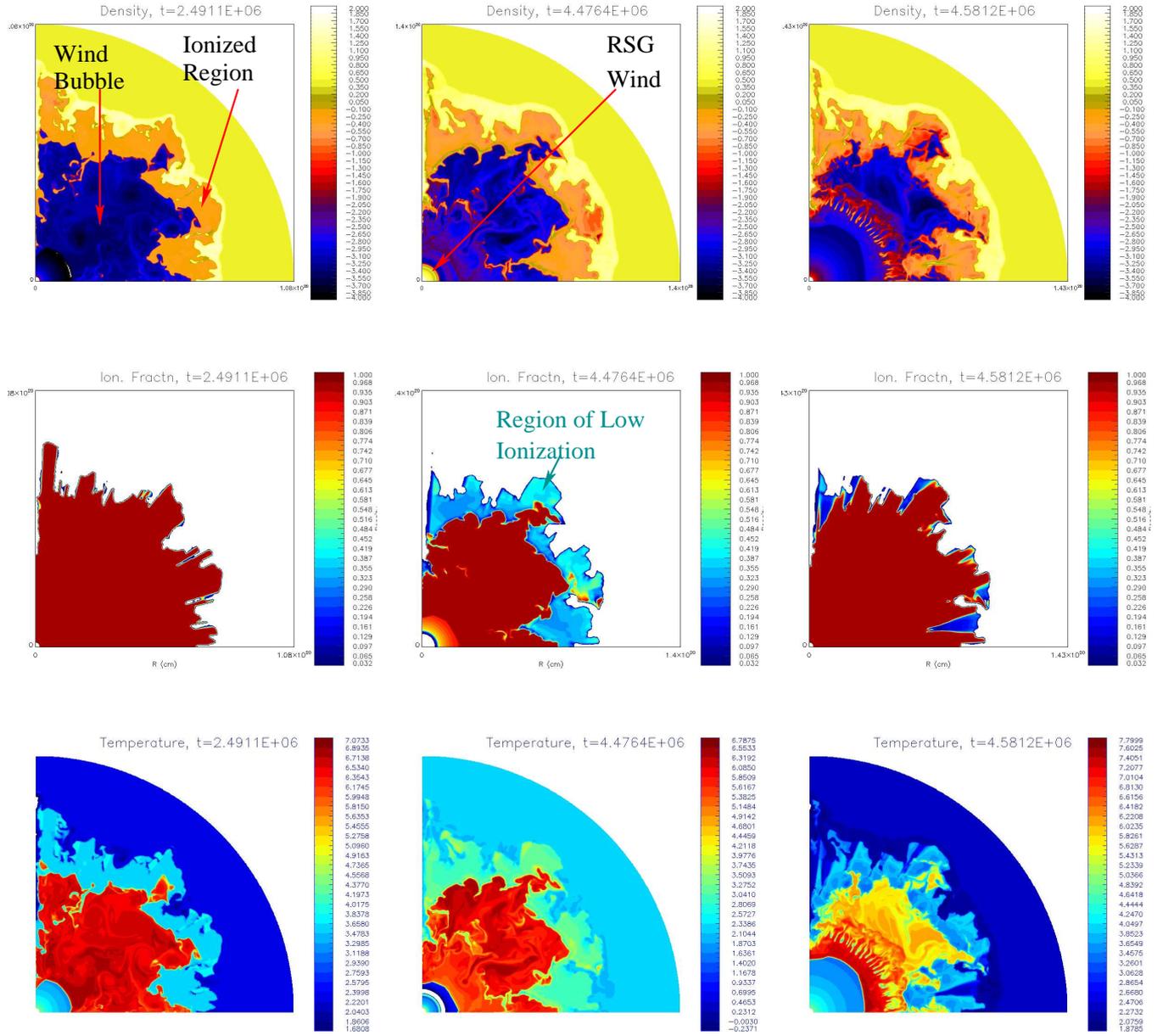}
\caption{Snapshots from a calculation of a wind bubble evolution
  around a 40 $\msun$ star (600 $\times$ 400 zones), computed with
  AVATAR using an expanding grid.  Density (top row), ionization
  fraction (middle row), and temperature (bottom row).  L
  $\Rightarrow$ R (a) The MS wind bubble (dark blue), expanding within
  an ionized region (orange) formed by photo-ionization. The
  ionization front is trapped in the dense shell (thin light yellow
  region), which is unstable to ionization front instabilities.  The
  wind bubble is hot, but the ionized region is only at $\sim$ 10$^4$
  K (b) The RSG shell within the MS bubble, a few parsecs from the
  star. In the RSG stage the star is cool, the ionization fraction
  decreases considerably, and the material in the bubble interior,
  starting at larger radii, begins to recombine. (c) The W-R wind,
  which breaks up the RSG shell, distributing its contents. Note the
  turbulent nature of the interior. The temperature slowly rises to
  X-ray producing levels again, but is still low in the outer
  parts. \label{fig:bubhydro}}
\end{figure}

Figure \ref{fig:bubhydro} shows the density, ionization and
temperature of the star during the various stages. The star starts
life as a main-sequence (MS) O star, and remains on the MS for about
4.3 million years, with a fast, radiatively-driven wind.  In this
phase almost the entire bubble is fully ionized
(Fig.~\ref{fig:bubhydro}, row 2 column 1).  An inhomogeneous pressure
and density distribution develops in the MS stage, accompanied by
vorticity deposition near the inner shock. The inclusion of
photo-ionization results in the formation of a dense, lower
temperature ($\sim 10^4$ K) region of ionized material (orange)
outside the wind bubble during the MS phase. The nebula is fully
ionized, and the ionization front is trapped in the dense shell. The
ionization front is unstable to ionization-driven instabilities,
reminiscent of Rayleigh-Taylor instabilities, which are also seen in
galactic ionization fronts \cite{wn08}. We also see finger-like
projections protruding from the ionized region into the interior,
which may be the forerunners of the photo-ionized pillars seen for
instance in the Eagle Nebula.

As the star moves off the MS into the red supergiant (RSG) stage, the
mass-loss rate increases sharply to almost 10$^{-4} \msun {\rm
  yr}^{-1}$, and the wind velocity drops to about 15 km s$^{-1}$.
Although much of the mass from the star is lost in this stage, the RSG
wind fills only a small fraction of the bubble.  The surface
temperature being much lower, the star is no longer able to ionize the
entire region. The ionizing radiation drops considerably and
recombination reduces the ionization fraction to $\sim$ 30\% in the
outer regions (Fig~\ref{fig:bubhydro}, row 2 column 2). As shown in
\cite{dwarkadas07c}, the RSG shell is also unstable. After about
200,000 years in the RSG stage, the star loses its outer envelope and
becomes a W-R star, with a fast wind and a mass-loss rate only
slightly lower than that in the RSG phase.  The high temperature of
the W-R star re-ionizes the entire region.  The higher momentum W-R
wind, due to its high velocity, collides with the dense RSG shell and
breaks it up, scattering the dense material into a turbulent W-R
nebula.

Turbulence within the bubble leads to the formation of vortices in the
interior of the bubble in the MS stage, driven by deposition of
vorticity at the reverse shock. We are aware that in this 2.5D
implementation, fluid that would normally flow across the boundary is
funneled along the boundary, leading to spurious jet-like structures
along the axes. We have discussed the hydrodynamics due to such
structures extensively in a purely hydrodynamical case
\cite{dwarkadas08}. We have employed methods such as jiggling the grid
in the azimuthal direction by half a grid zone every few timesteps to
reduce this noise. It is still present however, we have been unable to
eliminate it completely. Notwithstanding, as we have shown and
quantified in \cite{dwarkadas08}, there are physical reasons for
turbulence to be produced in a wind-bubble. Considerably more
turbulence is produced due to deposition of vorticity at the inner
shock, due to the various instabilities, and the onset of the W-R
wind, than due to the numerics. Vorticity deposition at the inner wind
termination shock, which is notably aspherical, produces much of the
turbulence in the MS stage. In the W-R phase, turbulence is mostly due
to the W-R wind interacting with, and breaking up, the RSG shell, the
resulting instabilities \cite{dwarkadas07c}, and mixing of the
material. The addition of stellar ionizing photons is the new
ingredient in this simulation, which results in ionization front
instabilities in the MS stage and therefore adds to the
turbulence. Thus while there is some uncertainty near the axes, the
turbulence in the bubble is quantifiably real, and not an artifact of
the numerics. It is accentuated by the 2D nature of the simulation,
since 2D turbulence results in an inverse cascade in energy
\cite{dwarkadas08}.  The turbulence decreases in the RSG phase.  The
collision of the W-R and RSG shells leads to a substantial increase in
turbulence in the final stages of the bubble, leading to a messy
structure into which any subsequent supernova shock wave will expand.

The temperature of the nebula during the evolution is shown in the
bottom row of Fig~\ref{fig:bubhydro}. Note how, in the early W-R
stage, the temperature is still low in the outer wind-blown
bubble. Thermal conduction has been suggested to reduce the
temperature (e.g.~Zhekov \& Park 2011), but our calculations do not
show any need for it.

How do our models compare to previous work? The gross features have
been outlined in many of the works mentioned in \S \ref{sec:intro},
and our models as expected do faithfully reproduce
them. Interestingly, our models reproduce multi-dimensional features
such as ionization front instabilities, seen in many galactic
ionization front models \cite{wn08} but not mentioned in previous
simulations of wind bubbles. The closest work to which our models
could be compared are those of \cite{arthur09, ta11}. Our 1D models
(not shown) agree quite well with their 1D simulations when run with
similar parameters, although our resolution of tens of thousands of
zones seems much higher.  The 2D models shown here are harder to
compare without having quantitative data. It is clear that we again
reproduce the basic features. Our shock fronts seem sharper and
instabilities far more pronounced, although it is not clear whether
this is a resolution effect, or merely due to the reproduction of the
figures in the papers from Arthur's group. One thing that our
simulations reveal clearly is the recombination of the material in the
red-supergiant stage as the surface temperature of the star
dramatically decreases, leading to a large decrease in the ionization
fraction. This is visible in some of their 1D simulations, but there
are no good figures clearly illustrating the evolution of the
ionization fraction in their multi-dimensional models. We feel that,
while all models reproduce the gross features, our work improves on
previous work in terms of capturing the various discontinuities,
elucidating the turbulence and instabilities, and understanding the
evolution of the ionization fraction.

\section{X-ray Emission:} Having computed the structure of the bubble, we 
proceed to compute the X-ray spectrum from the bubble at each
timestep.  We use the ISIS package \cite{houck2000} to compute the
X-ray emission. The {\tt vmekal} model in XSPEC, commonly used to fit
the observed X-ray spectra \cite{chuetal03a}, is used to model the
spectrum, although any other available XSPEC model can be
substituted. We have also tried the Raymond-Smith model ({\tt
  vraymond}), which gave similar results.

\begin{figure}[ht]
\includegraphics[scale=0.7]{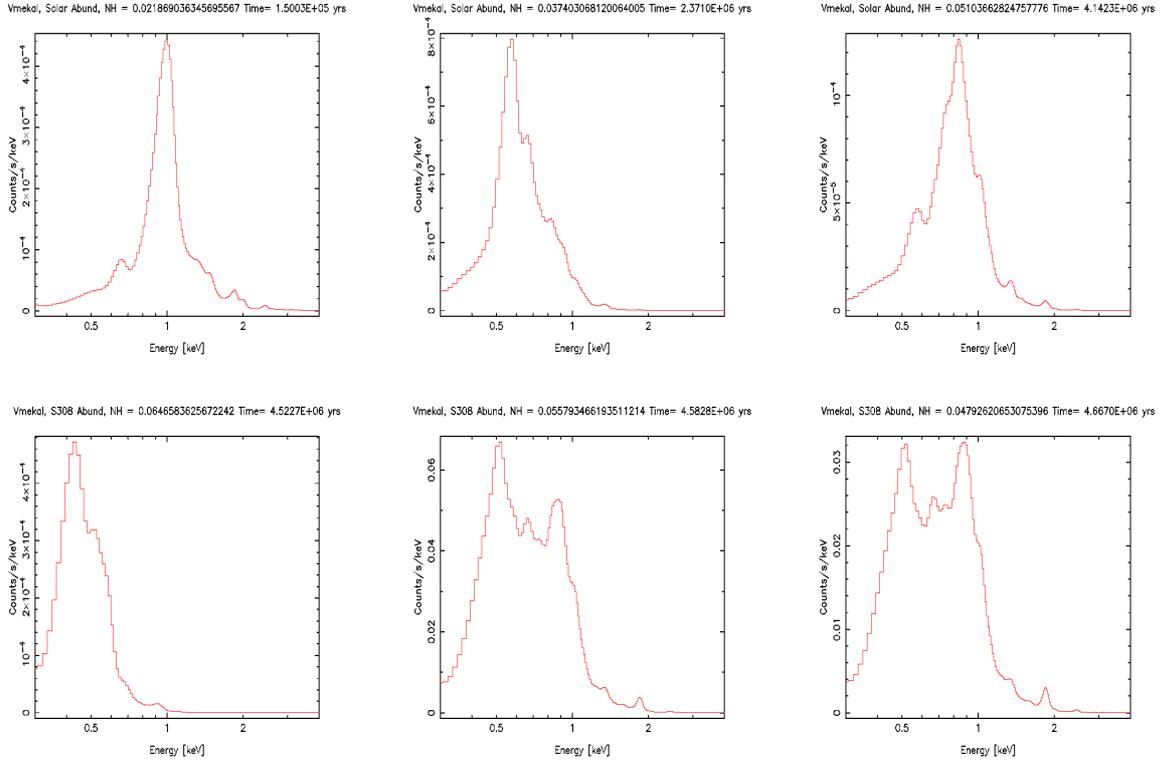}
\caption{\small (LHS) Simulated X-ray spectra for a point source
  calculated from simulations shown in Fig~\ref{fig:bubhydro}. The
  spectra were calculated from a single simulation. The first 3 are
  computed when the nebula is in the MS stage, the next 3 in the W-R
  stage. Solar abundances were assumed for the MS stage, while
  abundances characteristic of S308 were used for the W-R stage. The
  X-ray model used, the absorption column (in units of 10$^{22} {\rm
    cm}^{-2}$) and the Time in years is noted in the title of each
  plot. The scale is the count-rate that would be received for a
  source at 1.5 kpc distance in a 50 ks ACIS observation. The count
  rate is generally very low in our simulated spectra in the MS stage,
  about 10$^{-4}$ counts s$^{-1}$ keV$^{-1}$, and it is clear that
  such a bubble would not be visible in X-rays in the MS phase. It can
  rise by about 2 orders of magnitude or so towards the end of the W-R
  stage, which is still quite low as compared to the data on
  S308. This seems to be generally true over all our simulations, in
  agreement with the fact that diffuse X-ray emission is not generally
  observed. \label{fig:spectra} }
\end{figure}

The 2D data are read into ISIS. ISIS calls the appropriate XSPEC
routine (default: {\tt vmekal}) to calculate the spectrum in each
cell, using the values of various quantities such as temperature,
density, velocity in that cell derived from the hydrodynamic
simulations.  For the run presented here that means computing the
spectrum in each of 240,000 cells. The absorption column beyond that
cell is calculated, summed over all cells along the radius, and used
to calculate the absorption.  The spectra from all cells are then
added up, and convolved with the Chandra ACIS instrument response and
area functions, for a source distance of 1.5kpc distance, integrated
over 50,000 s (Fig~\ref{fig:spectra}).  Foreground absorption of 2
$\times$10$^{20}$ cm$^{-2}$ is added to absorption by neutral material
surrounding the bubble, which can be as high as 2 $\times$10$^{21}$
cm$^{-2}$ (total absorption column given on figure title is in units
of 10$^{22}$ cm$^{-2}$).  Line broadening is added based on the
underlying fluid velocity, and seems to adequately reproduce the
broadening in observed spectra.  Solar abundances \cite{ag89} are used
for the MS and RSG stages, while abundances in the W-R phase are from
\cite{chuetal03a} for the W-R bubble S308, one of two W-R bubbles to
be detected in diffuse X-rays.

Although no attempt has been made to fit to a specific bubble, the
resulting X-ray spectra in the Wolf-Rayet stage do show resemblance to
observed X-ray spectra given in \cite{chuetal03a, zp11}. The computed
spectra are quite soft, generally peaking at or below 1 keV. In this
they resemble those shown by \cite{ta11}, although our spectra show
better energy resolution. The spectra in \cite{ta11} appear to have
somewhat similar shape at certain epochs, but much higher
counts. Exact comparison is not possible because their spectra are
convolved with the instrument response for EPIC/pn on {\it
  XMM-newton}, and the integration time is not given. The emission
appears to arise primarily from the denser regions, including the
instabilities and resulting dense clumps that are formed. The volume
of the emitting region is reduced by the presence of the ionized
region surrounding it, and the temperatures are found to be lower due
to recombination in the RSG stage. Spectra are calculated in
ionization equilibrium. While we have recently developed the
capability to carry out non-equilibrium ionization calculations in the
case of 1D calculations \cite{ddb10, deweyetal12}, we are still in the
process of expanding this capability to multi-dimensional
calculations.  Such complexity does not seem to be required to
reproduce the observed X-ray spectra, especially given that much of
the material in the bubble is over 10$^5$ years old. It is possible
though that in the early Wolf-rayet stage, when the ionization front
due to the Wolf-Rayet star has not crossed the entire bubble, and the
W-R wind has not yet filled the bubble, the plasma may not be in
ionization equilibrium and more detailed calculations may be
needed. We plan to carry these out in future.

\section{Summary}
Using a code that employs a self-consistent method for computing the
effects of photo-ionization on circumstellar gas dynamics, we have
modeled the formation of wind-driven nebulae around massive
stars. Using various X-ray emission models, we have computed detailed
X-ray spectra of simulated wind-blown nebulae from a 40 $\msun$ star,
as would be seen with the ACIS instrument on Chandra. Comparing with
observed X-ray spectra, we find that at certain epochs of the
evolution, our synthetic spectra in the Wolf-Rayet (W-R) stage agree
quite well with those obtained from observed W-R nebulae. Unlike other
calculations, our detailed spectra indicate that diffuse X-ray
emission from most main sequence and W-R nebulae would {\em not} be
easily observable with currently available X-ray satellites, which is
consistent with the observational data \cite{chuetal03a,chuetal03b}.

In future we will present calculations for the wind medium around
stars of varying initial mass, taking stellar models with rotation and
magnetic field into account. We will also compute the X-ray emission
using different XSPEC models, study the locations from which the X-ray
emission arises, and produce maps showing the cells that are major
contributors to the X-ray emission.  These can be compared to density
and temperature maps to fully understand which regions are the major
contributors to the X-ray emission, and why the spectra are so soft.

\section{Acknowledgements}

VVD’s research was funded by grant TM9-0001X, provided by NASA through
the Chandra X-ray Observatory center, operated by SAO under NASA
contract NAS8-03060. DLR wishes to acknowledge sponsorship of this
work from an NSF cooperative agreement through the University
Corporation for Atmospheric Research on behalf of the National Center
for Atmospheric Research (NCAR). We are grateful to Dan Dewey for
patiently answering our questions and being super helpful in
general. We thank the anonymous referee for a critical report that
helped to substantially improve the focus of this paper and sharpen
the emphasis.






\begin{thebibliography}{xx}

\bibitem[Anders \& Grevesse(1989)]{ag89} Anders, E., \& Grevesse,
  N.\ 1989, \gca, 53, 197

\bibitem[Arthur(2007)]{arthur07} Arthur, S.~J.\ 2007, Revista Mexicana
  de Astronomia y Astrofisica Conference Series, 30, 64

\bibitem[Arthur(2009)]{arthur09} Arthur, S.~J.\ 2009, American 
Institute of Physics Conference Series, 1156, 285 

\bibitem[Chu et al.(2003)]{chuetal03b} Chu, Y.-H., Gruendl, R.~A., 
\& Guerrero, M.~A.\ 2003, Revista Mexicana de Astronomia y Astrofisica Conference Series, 15, 62 

\bibitem[Chu et al.(2003)]{chuetal03a} Chu, Y.-H., Guerrero, 
M.~A., Gruendl, R.~A., Garc{\'{\i}}a-Segura, G., 
\& Wendker, H.~J.\ 2003, \apj, 599, 1189 

\bibitem[Dewey et al.(2012)]{deweyetal12} Dewey, D., Dwarkadas, V.~V.,
  Haberl, F., Sturm, R., \& Canizares, C.~R.\ 2012, \apj, 752, 103

\bibitem[Dwarkadas(2005)]{dwarkadas2005} Dwarkadas, V.~V.\ 2005, 
\apj, 630, 892 

\bibitem[Dwarkadas(2007)]{dwarkadas07a} Dwarkadas, V.~V.\ 2007, \apss, 307, 153 

\bibitem[Dwarkadas(2007)]{dwarkadas07c} Dwarkadas, V.~V.\ 2007, 
\apj, 667, 226 

\bibitem[Dwarkadas(2008)]{dwarkadas08} Dwarkadas, V.~V.\ 2008, 
Physica Scripta Volume T, 132, 014024 

\bibitem[Dwarkadas et al.(2010)]{ddb10} Dwarkadas, V.~V., 
Dewey, D., \& Bauer, F.\ 2010, \mnras, 407, 812 

\bibitem[Frank \& Mellema(1994)]{fm94b} Frank, A., \& Mellema,
  G.\ 1994, \aap, 289, 937

\bibitem[Freyer et al.(2003)]{fhy03} Freyer, T., Hensler, G., 
\& Yorke, H.~W.\ 2003, \apj, 594, 888 

\bibitem[Freyer et al.(2006)]{fhy06} Freyer, T., Hensler, G., 
\& Yorke, H.~W.\ 2006, \apj, 638, 262 

\bibitem[Garcia-Segura et al.(1996)]{glm96} Garcia-Segura, G., Langer,
  N., \& Mac Low, M.-M.\ 1996, \aap, 316, 133

\bibitem[Garcia-Segura et al.(1996)]{gml96} Garcia-Segura, G., Mac
  Low, M.-M., \& Langer, N.\ 1996, \aap, 305, 229

\bibitem[Houck \& Denicola(2000)]{houck2000} Houck, J.~C., \&
  Denicola, L.~A.\ 2000, Astronomical Data Analysis Software and
  Systems IX, 216, 591


\bibitem[Rosenberg(1995)]{r95} Rosenberg, D.~L.\ 1995, Ph.D.~Thesis,


\bibitem[Toal{\'a} \& Arthur(2011)]{ta11} Toal{\'a}, J.~A., \& Arthur,
  S.~J.\ 2011, \apj, 737, 100

\bibitem[van Marle et al.(2011)]{vanmarleetal11} van Marle, A.~J.,
  Keppens, R., \& Meliani, Z.\ 2011, Bulletin de la Societe Royale des
  Sciences de Liege, 80, 310

\bibitem[van Marle et al.(2006)]{vanmarleetal06} van Marle, A.~J.,
  Langer, N., Achterberg, A., \& Garc{\'{\i}}a-Segura, G.\ 2006, \aap,
  460, 105

\bibitem[van Marle et al.(2005)]{vanmarleetal05} van Marle, A.~J.,
  Langer, N., \& Garc{\'{\i}}a-Segura, G.\ 2005, \aap, 444, 837

\bibitem[Weaver et al.(1977)]{weaver77} Weaver, R., McCray, R.,
  Castor, J., Shapiro, P., \& Moore, R.\ 1977, \apj, 218, 377

\bibitem[Whalen \& Norman(2008)]{wn08} Whalen, D.~J.,
  \& Norman, M.~L.\ 2008, \apj, 672, 287

\bibitem[Wrigge et al.(2005)]{wriggeetal2005} Wrigge, M., Chu, Y.-H., 
Magnier, E.~A., \& Wendker, H.~J.\ 2005, \apj, 633, 248 

\bibitem[Zhekov \& Park(2011)]{zp11} Zhekov, S.~A., \&
  Park, S.\ 2011, \apj, 728, 135

\end{thebibliography}







\end{document}